\documentclass{article}
\title{Scattering theory and discrete-time quantum walks}
\author{Edgar Feldman \\ Department of Mathematics \\ Graduate Center of the 
City University of New York \\ 365 Fifth Avenue \\ New York, NY 10016 \and
Mark Hillery \\ Department of Physics, Hunter College of CUNY \\
695 Park Avenue \\ New York, NY 10021}
\begin{document}
\maketitle
\begin{abstract}
We study quantum walks on general graphs from the point of view of scattering
theory.  For a general finite graph we choose two vertices and attach one half 
line to each, and consider walks that proceed from one half line, 
through the graph, to the other.  The probability of starting on one line 
and reaching the other after $n$ steps can be expressed in terms of the 
transmission amplitude for the graph.
\end{abstract}

Classical random walks on graphs can be used to construct algorithms 
that solve 2-SAT, graph connectivity problems, and for finding satsifying 
assignments for Boolean functions.  A hope is that recently defined
quantum walks will prove similarly useful in the development of quantum  
algorithms.  In fact, there has been consderable recent progress is this
regard.  It has been shown that it is possible to
use a quantum walk to perform a search on the hypercube faster than
can be done classically \cite{shenvi}.  In this problem the number of
steps drops from $N$, which is the number of vertices, in the classical
case to $\sqrt{N}$ in the quantum case.  A much more dramatic 
improvement has recently been obtained by Childs, \emph{et al.} \cite{childs2}.
They constructed an oracle problem that can be solved by a quantum
algorithm based on a quantum walk exponentially faster than is possible
with any classical algorithm.  Quantum algorithms that are faster than any
classical one have been found for for searching databases laid out in $D$ 
dimensions using a continuous time walk \cite{childs4} and in two dimensions
using a discrete time walk \cite{ambainis1}.  Quantum-walk algorithms have 
also been found for element distinctness \cite{ambainis2}, finding triangles
in graphs \cite{magniez}, subset finding \cite{childs5}, and determining 
whether a set of marked elements, which is promised to be of a certain
size, exists or not \cite{szegedy}. 

There are a number of different kinds of quantum walks two of which are
discussed in the recent review by Kempe \cite{kemperev}.  The type
of walk we shall discuss here is based on thinking about the graph 
as an interferometer with optical multiports as the nodes 
\cite{hillery}.  In this case the walk takes place on the edges of the
graph, rather than the vertices, and each edge has two states, one
corresponding to traversing the edge in one direction and the other to
traversing the edge in the opposite direction.  

Several proposals have been made for the physical realization of quantum
walks \cite{travaglione}-\cite{knight2}.  The ones closest in spirit to the
walks studied here are the realizations that employ optical methods,
either linear optical elements \cite{zhao,jeong} or cavities 
\cite{knight1,knight2}.  The last two references show that an experimental
quantum walk has, in fact, been carried out, though it was not intrepreted
as such at the time \cite{bouwmeester}.

Here we wish to study the connections between quantum walks and scattering
theory.  We begin with a graph, choose two vertices, and attach two 
semi-infinite lines, which we shall call tails, to it, one to each of the 
selected vertices.  Each of the half-lines is made up of vertices and edges,
and the particle propagates freely on the tails, that is the particle 
simply advances one step along the tail with each time step of the walk.  
Thus we can start the walk on one of the tails, have it progress into the
original graph, and emerge onto the opposite tail.  This type of 
arrangement allows us to define an S matrix for the original graph, with
the amplitude to get from one tail to the other being called the 
transmission amplitude.  As we shall see properties of quantum walks starting 
on one tail and ending on the other can be expressed in terms of the 
transmission amplitude of the graph.

The application of scattering theory to quantum walks was first done by
Farhi and Gutmann \cite{farhi}.  They studied the propagation of 
continuous-time walks through trees, and were able to turn it into a
scattering problem by attaching semi-infinite tails to the trees. By
doing so they were able to place bounds on the time necessary to 
go from the root of the tree to one of the leaves.

We begin by defining the type of quantum walk that will be used throughout
this paper.  It was originally presented in \cite{hillery}.  We imagine
a particle on an edge of a graph; it is this particle that will make the
walk.  Each edge has two states, one going in one direction, the other
going in the other direction.  That is, if our edge is between the
vertices $A$ and $B$, which we shall denote as $(A,B)$, it has two
orthogonal states, $|A,B\rangle$, corresponding to the particle being
on $(A,B)$ and going from $A$ to $B$, and $|B,A\rangle$, corresponding to 
the particle being on $(A,B)$ and going from $B$ to $A$.  The collection
of all of these edge states is a basis for a Hilbert space, and the states
of the particle making the walk lie in this space.

Now that we have our state space, we need a unitary operator that advances
the walk one step.  Let us first consider how this works for a walk
on the line.  We shall label the vertices by the integers.  In this case,
the states of the system are $|j,k\rangle$, where $k=j\pm 1$. 
The vertices can be thought of as scattering centers.  Consider what happens
when a particle, moving in one dimension, hits a scattering
center.  It has a certain amplitude to continue in
the direction it was going, i.e.\ to be transmitted, and an amplitude
to change its direction, i.e.\ to be reflected.  The scatterer
has two input states, the particle can enter from either the
right or the left, and two output states, the particle can leave heading
either right or left.  The scattering center defines a unitary transformation
between the input and output states.

This gives us the transition rules for our
quantum walk.  Suppose we are in the state $|j-1,j\rangle$.  If the
particle is transmitted it will be in the state $|j,j+1\rangle$, and if
reflected in the state $|j,j-1\rangle$.  Let the transmission amplitude
be $t$, and the reflection amplitude be $r$.  We then have the
transition rule
\begin{equation}
|j-1,j\rangle\rightarrow t|j,j+1\rangle +r|j,j-1\rangle ,
\end{equation}
where unitarity implies that $|t|^{2}+|r|^{2}=1$.
The other possibility is that the particle is incident on vertex
$j$ from the right, that is it is in the state $|j+1,j\rangle$.
If it is transmitted it is in the state $|j,j-1\rangle$, and if
it is reflected, it is in the state $|j,j+1\rangle$.  Unitarity
of the scattering transformation then gives us that
\begin{equation}
|j+1,j\rangle\rightarrow t^{\ast}|j,j-1\rangle -r^{\ast}|j,j+1\rangle .
\end{equation}
These rules specify our walk.

The case $t=1$ and $r=0$ corresponds to free particle propagation; a
particle in the state $|j,j+1\rangle$ simply moves one step to the
right with each time step in the walk.  If $r\neq 0$, then there is
some amplitude to move both to the right and to the left.  

So far we have only considered vertices at which two edges meet, but
if we are to construct graphs more complicated than lines, we need
to see how a vertex with more that two edges emanating from it behaves.
If a vertex treats all edges entering it in an equivalent fashion, then we
have a particulary simple situation, because the edges of the graph do not
have to be labelled, and this is the situation we shall consider.
Let the vertex at which
all of the edges meet be labelled by $O$, and the opposite ends of
the edges be labelled by the numbers $1$ through $n$.  For any 
input state, $|kO\rangle$, where $k$ is an integer between $1$ and
$n$, the transition rule is that the amplitude to go the output
state $|Ok\rangle$ is $r$, and the amplitude to go to any other
output state is $t$.  That is, the amplitude to be reflected is $r$,
and the amplitude to be transmitted through any of the other edges
is $t$.  Unitarity places two conditions on these amplitudes
\begin{eqnarray}
(n-1)|t|^{2}+|r|^{2}=1 \nonumber \\
(n-2)|t|^{2}+r^{\ast}t+t^{\ast}r =0 .
\end{eqnarray}
As an example, for the case $n=3$, possible values of $r$ and $t$
are $r=-1/3$ and $t=2/3$.  
 
In order to construct a walk for a general graph, one chooses a unitary
operator for each vertex, i.e.\ one that maps the states coming into
a vertex to states leaving the same vertex.  One step of the walk consists
of the combined effect of all of these operations; the overall unitary
operator, $U$, that advances the walk one step is constructed from the local
operators for each vertex.  Explicitly, the edge state $|AB\rangle$,
which is the state for the particle going from vertex $A$ to vertex $B$, will
go to the state $U_{B}|AB\rangle$ after one step, where $U_{B}$ is
the operator corresponding to vertex $B$.  This prescription guarantees
that the overall operation is unitary.

\begin{picture}(300,100)
\put(50,50){\line(1,0){30}}
\put(80,50){\circle*{5}}
\put(80,50){\line(1,0){30}}
\put(110,50){\circle*{5}}
\put(110,50){\line(1,1){25}}
\put(110,50){\line(1,-1){25}}
\put(135,75){\circle*{5}}
\put(135,25){\circle*{5}}
\put(135,75){\line(1,-1){25}}
\put(135,25){\line(1,1){25}}
\put(160,50){\circle*{5}}
\put(160,50){\line(1,0){30}}
\put(190,50){\circle*{5}}
\put(190,50){\line(1,0){30}}
\end{picture}
\newline
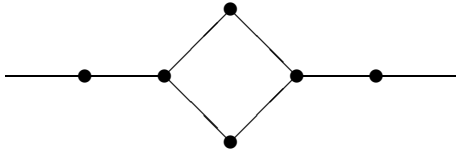
Figure 1: Graph consisisting of diamond-shaped region where scattering
occurs, and two tails on which the particle propagates freely.
\newline \newline

Let us put all of this together in a very simple example.  Consider the
graph in Figure 1, 
where each of the vertices where two edges meet have $t=1$ and $r=0$,
while the three-edge vertices are of the type discussed 
previously, with $r=-1/3$ and $t=2/3$. The graph
goes to negative infinity on the left and plus infinity on the right.
To fit this into the framework discussed previously, we can consider
the diamond shaped region consisting of the four edges between the
two vertices with three edges meeting at them as the original graph 
and the two lines going to plus and minus infinity as the tails. 

We can find the unnormalized eigenstates for this graph, and one set of
them can be described as having an incoming wave from the left, an
outgoing transmitted wave going to the right, and a reflected wave
going to the left. A second set will have an incoming wave from the right,
an outgoing trasmitted wave to the left, and a refelected wave to the
right.  Finally, there may be bound states, i.e.\ eigenstates that are
localized in the region between the two vertices with three edges.
Now, let us denote the left three-edge
vertex by $0$ and the right one by $2$, and number the vertices on the
lines correspondingly, from $2$ to plus infinity to the right and
from $0$ to minus infinity to the left.  The eigenstates with a wave
incident from the left take the form
\begin{eqnarray}
|\Psi\rangle & = &  \sum_{j=-\infty}^{-1}(e^{ij\theta}|j,j+1\rangle +
r(\theta )e^{-i(j+1)\theta}|j+1,j\rangle )+|\Psi_{02}\rangle \nonumber \\
 & &+\sum_{j=2}^{\infty}t(\theta )e^{i(j-2)\theta}|j,j+1\rangle ,
\end{eqnarray} 
where $|\Psi_{02}\rangle$ is the part of the eigenfunction between 
vertices $0$ and $2$, and $e^{-i\theta}$ is the eigenvalue of the
operator $U$ that advances the walk one step.  The first term can be thought
of as the incoming wave; it is confined to the region between negative
infinity and $0$, and consists of states in which the particle is moving
to the right.  The term proportional to $r(\theta )$ is the reflected wave.
It is also confined to the region between negative infinity and $0$, but
consists of states in which the particle is moving to the left.  Finally,
the term proportional to $t(\theta )$ is the transmitted wave, 
being confined to
the region from $2$ to infinity, and consisting of states with the particle
moving to the right.  Inserting the above expression into the equation
$U|\Psi\rangle =e^{-i\theta}|\Psi\rangle$ we find
\begin{equation}
t(\theta )=\frac{8e^{3i\theta}}{9-e^{4i\theta}} .
\end{equation}

Suppose we start the quantum walk in the state $|-1,0\rangle$, and after
each time step measure the state to determine whether the particle is in
the state $|2,3\rangle$.  Denote the probablity that we find the particle
there after $n$ steps after not having found it there on the previous
$n-1$ steps by $q(n)$.  This probability can be expressed in terms of the
transmission amplitude \cite{hillery2}.  Setting $z=e^{i\theta}$ we
can analytically extend the transmission amplitude to the complex
plane by setting
\begin{equation}
t(z)=\frac{8z^{3}}{9-z^{4}} .
\end{equation}  
We find that
\begin{equation}
\label{qn}
q(n)=\left|\frac{1}{n!}\left.\frac{d^{n}}{dz^{n}}t(z)\right|_{z=0}
\right|^{2},
\end{equation}
that is, the square of the magnitude of the coefficient of $z^{n}$ in
the Taylor series expansion of $t(z)$ about the point $z=0$.
The probability that we first find the particle on the edge 
$|2,3\rangle$ after any number of steps, $P_{out}$, is given by
\begin{equation}
\label{Pout}
P_{out}=\sum_{n=1}^{\infty}q(n)=\frac{1}{2\pi}\int_{0}^{2\pi} d\theta\,
|t(\theta )|^{2} .
\end{equation}
These results are general, they hold if the diamond-shaped region is
replaced by any finite graph, $G$, where $t(z)$ is the transmission 
amplitude for $G$  \cite{hillery2}.  In the case of this
particular graph we find that
\begin{eqnarray}
q(n)& =& \left\{ \begin{array}{cr}\left(\frac{8}{9^{(n+1)/4}}\right)^{2} & 
{\rm if}\ n=3\ {\rm mod}4 \\ 0 & {\rm otherwise} \end{array} \right. 
\nonumber \\
P_{out}& = & \frac{4}{5} .
\end{eqnarray}
For small $n$, it is relatively straightforward to verify the expression 
for $q(n)$ by adding up the amplitudes for the possible paths from 
$|-1,0\rangle$ to $|2,3\rangle$ of length $n$.

The connection between the transmission coefficent and quantum walk 
probabilities discussed in the previous paragraph is completely general,
i.e.\ it holds for any graph.  For any graph, $G$ we can define transmission
and reflection coefficients, these can be extended analytically to a
region of the complex plane that includes the unit disc, and Eqs.\ (\ref{qn})
and (\ref{Pout}) hold \cite{hillery2}.  

Other aspects of the quantum walk on this graph can be illuminated by viewing
it as a scattering process.  For example, this graph has bound states, that
is eigenstates of $U$ whose support is confined to the diamond-shaped
region.  In fact, it has four of them, one for each of the eigenvalues
$\pm 1$, and $\pm i$ \cite{hillery2}.  The state resulting from a quantum 
walk starting on one of the tails must be orthogonal to the bound states 
at all times.  This is again a general feature of quantum walks that holds
for any graph.  If the graph, $G$, has bound states, i.e.\ eigenstates of 
$U$, whose support is confined to $G$ (and, consequently, have no 
overlap with the tails), the state of a quantum walk starting on one of the 
tails will be orthogonal to all of the bound states at all times.  This
could place limits on the properties of the graph that a quantum walk could
sample. 

We have seen that the properties of a walk starting on the 
left-hand tail can be described by the transition amplitude to go through
the diamond from left to right.  Similarly, the properties of a walk starting 
on the right-hand tail are described by the transmission amplitude to go 
through the diamond from right to left.  These two amplitudes will be
identical if the quantum walk obeys time-reversal invariance, which this
particular walk does \cite{hillery2}.  The time reversal operator for a
quantum walk, $\hat{T}$, is an anti-unitary operator whose action on the
edge state $|A,B\rangle$ is given by
\begin{equation}
\hat{T}|A,B\rangle = |B,A\rangle .
\end{equation}
A quantum walk is time-reversal invariant if $\hat{T}U\hat{T}=U^{-1}$,
that is, the operator that moves the walk forward one step, when
conjugated with the time-reversal operator, becomes the operator that takes
the walk back one step.  The example given here obeys this condition.  A
quantum walk on a general graph $G$, which satisfies this condition, will
have its two transmission amplitudes, one describing transmission of a 
particle coming from the right and the other transmission of a particle
coming from the right, equal \cite{hillery2}.  

We have shown that that there is a close connection between quantum walks 
on a general graph and the S matrix of that graph. In particular, 
the probabilities that describe the walk can be expressed in terms of the 
reflection and transmission amplitudes, or their analytic extensions, 
of the graph on which the walk is taking place.  Many of the results in
the classical theory of random walks on graphs can be proven by exploiting
the connection between these walks and the theory of electrical networks
\cite{bollobas}.  The results here suggest that in the case of quantum
walks, it may be possible to exploit quantum scattering theory to prove
results about quantum walks.  If quantum walks are to be used as a basis
for quantum algorithms, a better understanding of their properties is
necessary.  The application of methods from quantum scattering theory
may be able to help us gain this understanding.  

This research of one of the authors (M.\  H.) was supported by the National 
Science Foundation under grant PHY 0139692.

\bibliographystyle{unsrt}

\end{document}